\documentclass[%
 aip,
 pof,
 amsmath,amssymb,
preprint,%
]{revtex4-1}

\usepackage{graphicx}
\usepackage{dcolumn}
\usepackage{bm}
\usepackage[mathlines]{lineno}
\linenumbers\relax 
\usepackage{wasysym}
\usepackage[utf8]{inputenc}
\usepackage[T1]{fontenc}
\usepackage{etoolbox}
\usepackage{siunitx}
\usepackage{xcolor}
\usepackage{pifont}
\usepackage{hyperref}
\graphicspath{{./pic/}}

\usepackage[draft]{changes}
\usepackage{soul}
\makeatletter
\def\@email#1#2{%
 \endgroup
 \patchcmd{\titleblock@produce}
  {\frontmatter@RRAPformat}
  {\frontmatter@RRAPformat{\produce@RRAP{*#1\href{mailto:#2}{#2}}}\frontmatter@RRAPformat}
  {}{}
}%
\makeatother

\usepackage{CJK}

\begin{document}
\preprint{AIP/123-QED}

\nolinenumbers

\begin{CJK*}{UTF8}{gbsn}

\title{Cascades and Kolmogorov's lognormal scaling in two-dimensional
bacterial turbulence}
\author{Yongxiang Huang (黄永祥)}
\email{yongxianghuang@\{gmail.com,xmu.edu.cn\}}
\affiliation{State Key Laboratory of Marine Environmental Science \& Center for Marine Meteorology and Climate Change \&  College
of Ocean and Earth Sciences,
Xiamen University, Xiamen,  China}

\affiliation{Fujian Engineering Research Center for Ocean Remote Sensing
Big Data, Xiamen, China}

\date{\today}

\begin{abstract}

Collective movements of bacteria exhibit a remarkable pattern of turbulence-like vortices, in which the Richardson cascade plays an important role. In this work, we examine the energy and enstrophy cascades and their associated lognormal statistics using experimental velocity field data. The coherent structure observed on a large scale is due to the presence of the inverse energy cascade; while the kinetic energy is dissipated at all scales, since these active movements occur below the fluid viscosity scale. The forward enstrophy cascade occurs with injection at all scales and may be represented by other nonlinear interactions that are not captured by the existing experimental data.
Furthermore, the lognormal statistics for both energy dissipation and enstrophy fields are verified in accordance with the Kolmogorov 1962 refined theory of turbulence. Their scaling exponents can be well described by the lognormal formula with intermittency parameters comparable with those of the three-dimensional hydrodynamic turbulence. The joint analysis of the multifractal measures of the energy dissipation rate and enstrophy follows an ellipse model from the lognormal statistics. Our results confirm the coexistence of the inverse energy cascade and the intermittency correction of the velocity scaling in this active fluid system. An inverse energy cascade diagram below the fluid viscosity is summarized to describe the observed two-dimensional bacterial turbulence. Our work provides an example of an active-flow model benchmark.
\end{abstract}

\maketitle
\end{CJK*}


 \section{Introduction}

One hundred years ago, Lewis Fry Richardson proposed his celebrated cascade to describe the movement of turbulent flows in which the kinetic energy (i.e., square of velocity) is transferred from the large-scale to the
small-scale vortexes until the smallest one, known as the dissipation scale, converts
into heat. \citep{Richardson1922} It was then Kolmogorov who proposed his acclaimed
three-dimensional (3D) homogeneous and isotropic turbulence (HIT) theory to quantitatively describe this cascade idea
in 1941. \citep{Kolmogorov1941,Frisch1995}  To cope with the two-dimensional (2D) situation,
\citet{Kraichnan1967PoF} generalized the idea of the forward energy cascade so as the inverse energy cascade that 
the kinetic energy or other physical quantities could be transferred from the small-scale  to the large-scale vortexes.\citep{Batchelor1969PoF} 
The concept of cascade is now widely accepted to describe turbulent
flows or turbulence-like systems \citep{Young2017NaturePhysics,Jian2019PRE}
and has been treated as the cornerstone of turbulent models,
\citep{Chou1940CJP,Chou1945QAM,Landau1987,Pope2000,Davidson2004book} the
global circulation model of the atmosphere
and oceans, \citep{Thorpe2005book,Vallis2017Book,Lovejoy2019Book} to name a few.

\begin{figure*}[!htb]
\centering
 \includegraphics[width=0.65\linewidth,clip]{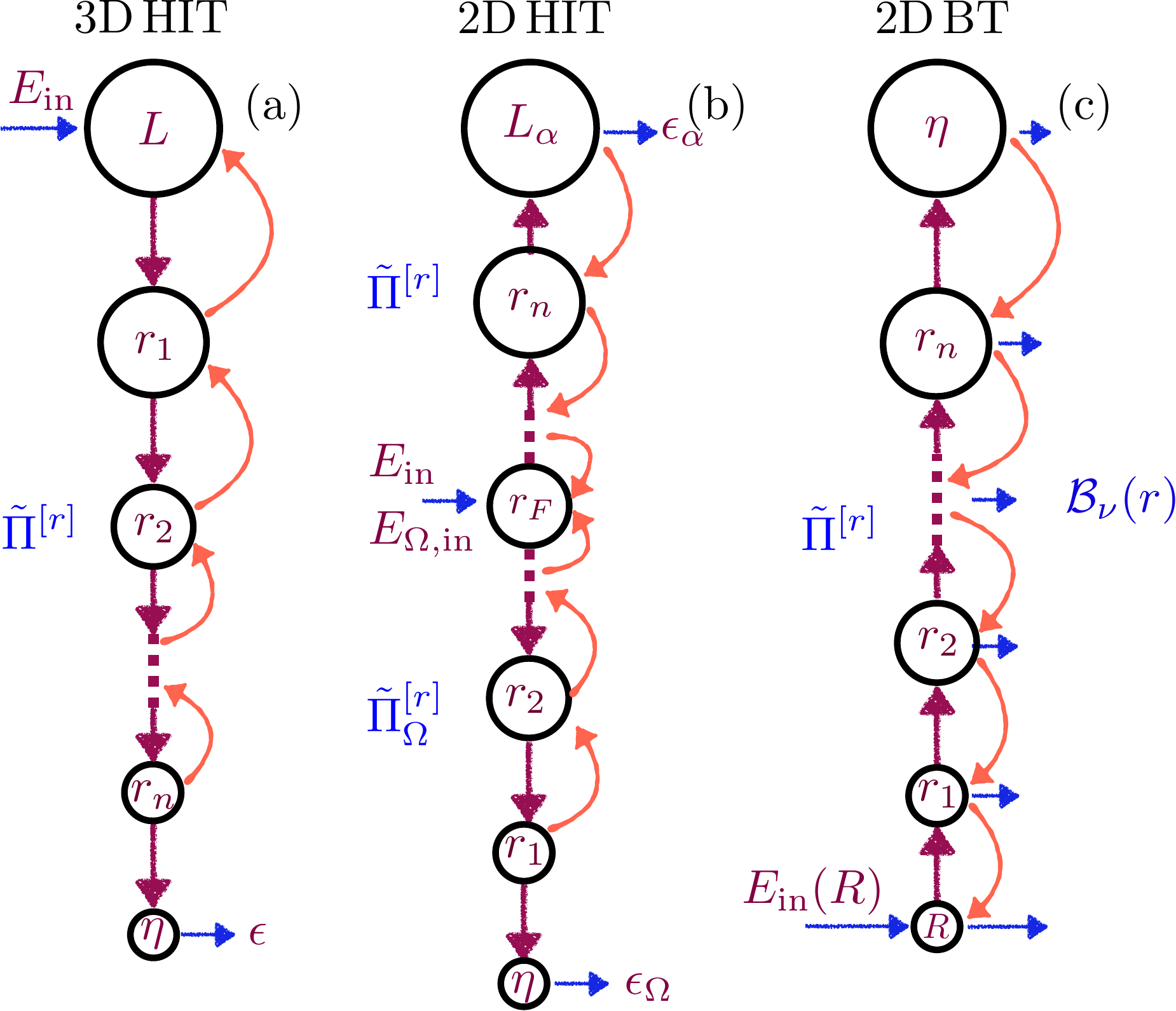}
  \caption{(Color online) (a)  Illustration of the forward energy cascade in the 3D homogeneous and isotropic turbulence, where the downward arrow indicates the direction of the mean forward cascade, and the curved arrow indicates the instantaneous inverse energy cascade. (b) Forward energy and inverse enstrophy cascades in 2D turbulence. 
  (c) The inverse energy cascade for the 2D bacterial turbulence that is below the fluid viscosity scale. The horizontal arrow implies the energy dissipated density function $\mathcal{B}_{\nu}(r)$ on scale $r$.
  }\label{fig:cascade}
\end{figure*}

To be a cascade, taking the 3D HIT as example, the kinetic energy is injected into the system at large scale $L$ with a rate $E_{\mathrm{in}}$, it is then forward transformed hierarchically from $L$ to small scale $r_1$, then to $r_2$, and so on to $\eta$,
the so-called dissipation scale; see an illustration in Fig.\,\ref{fig:cascade}\,(a). The corresponding energy transfer rates between scales are 
$\tilde{\Pi}^{[r_1]}$, $\tilde{\Pi}^{[r_2]},\cdots $, and so on, where $\tilde{\cdot}$ means the average of the ensemble. 
Assuming homogeneity and statistically stationary state over time, energy conservation requires the global balance between injection and dissipation,\citep{Frisch1995} which can generally be written as,
\begin{equation}
   E_{\mathrm{in}}= \int \mathcal{E}_{\mathrm{in}}(r) \mathrm{d}r =\int \mathcal{B}_{\nu}(r) \mathrm{d}r =\epsilon, \label{eq:balance}
\end{equation}
where $\mathcal{E}_{\mathrm{in}}(r)$  and $\mathcal{B}_{\nu}(r)$ are the energy injection rate density function due to the external forcing and the energy dissipation rate density function due to the fluid viscosity on scale $r$, respectively. Here, we consider only the cases $\mathcal{E}_{\mathrm{in}}(r)\ge 0$ and $\mathcal{B}_{\nu}(r)\ge 0$.
If there is a large-scale separation, that is a large Reynolds number $\mathrm{Re}=UL/\nu$, the ratio between the inertial force and the
viscosity force, where  $U$,  $L$  and  $\nu$ are the characteristic velocity,  spatial scale and  fluid viscosity, the effect of viscosity can be ignored for those mediate scales, that is $\eta \ll r\ll L$, known as  the inertial 
range. Thus, an asymptotic conservation relation $E_{\mathrm{in}}\simeq \tilde{\Pi}^{[r]}\simeq \epsilon$ is expected in the inertial range. Within this inertial range, the Kolmogorov  theory of 3D HIT then predicts  the Fourier power spectrum of the kinetic energy as $E(k)\propto k^{-5/3}$ (resp. $k=1/r$ is 
wavenumber).\citep{Kolmogorov1941,Frisch1995,Schmitt2016Book} This power law scaling prediction and the forward energy cascade 
are widely verified experimentally and numerically. \citep{Frisch1995}  A diagram of this classical forward cascade is shown in 
Fig.\,\ref{fig:cascade}\,(a), where the downward arrow indicates the direction of the mean forward energy cascade and the curved arrow 
indicates the inverse transfer of the kinetic energy. As pointed out by \citet{Lumley1992}, a real cascade is always bidirection; for 
example, both forward and inverse energy cascades coexist and cross different scales; see Fig.\,8 in Ref.\onlinecite{Lumley1992}.  The 
direction of the final cascade is thus the competition between the forward and inverse processes. Moreover, due to the intermittent 
distribution of the energy dissipation field, the scaling behavior of the velocity field deviates noticeably from the prediction of the 
Kolmogorov 1941 theory, known as intermittency.\citep{Kolmogorov1962,Frisch1995} The intermittency phenomenon has been interpreted in 
the framework of multifractality of the energy dissipation field,\citep{Benzi1984,Parisi1985,Meneveau1991JFM} and verified widely by experiments and numerical simulations.\citep{Stolovitzky1994,Chen1997PRL,Chen1997PRLb,Benzi2009PRE,Yu2010PRL,Huang2014JFM} Note that, in this conventional view of the 3D cascade, a large-scale separation ratio is required. Or, in other words, to have a cascade and the observed scaling behavior of the velocity field, there should be a wide range of fluid structures to interact with each other, which is often characterized by a large Reynolds number $\mathrm{Re}$. Therefore, the cascade and the associated intermittent behavior of the velocity field are often considered one of the main properties of high Reynolds number flows; \citep{Tsinober2009book} see a full discussion in Refs.\,\onlinecite{Alexakis2018PR,Zhou2021PR,Benzi2023PR}.

Concerning the 2D case, it does not make the problem easier because of the reduction of dimensionality. For example, as conceptualized by \citet{Kraichnan1967PoF}, if the injection of energy is on the intermediate scale, that is $r_F$, addition to energy conservation, the conservation of enstrophy (i.e. the square of vorticity) is also expected.\citep{Batchelor1969PoF} More precisely, the inverse energy cascade is expected when $r_F\ll r \ll L_{\alpha}$ with $E(k)\propto k^{-5/3}$, while the forward enstrophy cascade is expected when $\eta \ll r\ll r_F$ with $E(k)\propto k^{-3}$, where $L_{\alpha}$ is the Ekmann friction scale due to Ekmann friction;\citep{Boffetta2012ARFM} see Fig.\,\ref{fig:cascade}\,(b) for an illustration. In analogy to the 3D HIT case, the conservation law implies the following asymptotic relation $E_{\mathrm{in}}(r_F)\simeq -\tilde{\Pi}^{[r]}=\epsilon_{\alpha}$ for the inverse energy cascade and
$E_{\Omega,\mathrm{in}}(r_F)\simeq \tilde{\Pi}_{\Omega}^{[r]}=\epsilon_{\Omega}$ for the forward enstrophy cascade.
The experimental results confirm the existence of such dual cascades. \citep{Kraichnan1980RepProgPhys,Paret1999PRL,Chen2003PRL,Chen2006PRL,Alexakis2006PLA,Boffetta2007JFM,Xia2008PRL,Boffetta2010PRE,Falkovich2011PRE,Tan2014PoF,Fang2016PRL,Gauthier2019Science,Johnstone2019Science}
Unlike 3D hydrodynamic turbulence, there is no intermittent correction in the 2D inverse energy cascade.\citep{Falkovich1994PRE,Tsang2005PRE,Falkovich2011PRE,Tan2014PoF,Paret1999PRL}

It is interesting to note that the turbulence-like dynamics was also observed for several low Reynolds number flows; for example, the elastic turbulence with a typical Reynolds number $\mathrm{Re}=\mathcal{O}(0.1)$, \citep{Groisman2000Nature,Groisman2001Nature,Steinberg2021ARFM} the collective motion of a high concentration of bacteria with $\mathrm{Re}=\mathcal{O}(10^{-5})$ (also known as mesoscale turbulence, or active turbulence), \citep{Zhang2010PNAS,Wensink2012PNAS,Dunkel2013PRL,Koch2021PRL,Bratanov2015PNAS,Wu2017Science,Huang2017PoF,Chen2017Nature,Peng2021SA} and the lithosphere deformation with  $\mathrm{Re}=\mathcal{O}(10^{-24})$,\citep{Jian2019PRE} to name a few. The cascades in these mentioned systems, if they exist, should be more complex, as their external forces and dissipation mechanisms could be very different from those of the 3D and 2D HITs.
 For example, in bacterial turbulence, a remarkable
coherent flow pattern with spatial size greater than 10 times their mean body size $\tilde{R}$ was reported. \citep{Wensink2012PNAS,Bratanov2015PNAS} This large-scale
structure pattern is believed to be a consequence of an inverse energy cascade
through hydrodynamic interactions, in which kinetic energy is injected into the flow system through the stirring of bacteria.\citep{Bratanov2015PNAS,Doostmohammadi2017NC,Wang2017PRE,Linkmann2019PRL}  
Unlike conventional
turbulent flows with high Reynolds numbers, where inertia dominates on a wide
range of scales, the cascade of the bacterial turbulence, if it exists, is still considered below the viscosity scale. \citep{Wang2017PRE}
\footnote{The classical definition of 
the Kolmogorov scale or viscosity scale is written as $\eta=(\nu^3/\epsilon)^{1/4}$,
where $\epsilon$ is the energy 
dissipation rate. A typical value of $\eta$ in the ocean is around $0.2\sim
2\,\si{mm}$. In a turbulent pipe flow with a high Reynolds number with velocity, $1.8\,\si{ms^{-1}}$, $\eta$ is found
to be $25\,\si{\mu m}$. In the current system, the Reynolds number is nearly
zero, that is, $\mathrm{Re}\simeq \mathcal{O}(10^{-5})$. Therefore, it is reasonable to assume that the viscosity scale is higher than $50\,\si{\mu
m}$.}
There is strong competition between fluid inertia and viscosity. For example, this work will show that the kinetic energy is rapidly dissipated on all scales due to the effect of fluid viscosity; see the illustration in Fig.\,\ref{fig:cascade}\,(c), and more discussion in section\,\ref{sec:dis_cascade}. Thus,  cascades play an important role in the formation of this special active-fluid system.

Since its discovery,\citep{Wu2000PRL,Dombrowski2004,Zhang2010PNAS,Ishikawa2011PRL,Sokolov2012PRL,Wensink2012PNAS} the bacterial turbulence and its associated models have attracted more and more attention.\citep{Koch2021PRL,Wu2017Science,Kokot2017PNAS,Guillamat2017NatComm,Martinez2019NatPhys,Vcopar2019PRX,Opathalage2019PNAS,Shaebani2020NRP,Bar2020ARCM,Alert2020NatPhys,Koch2021PRL,Alert2022ARCMP,Aranson2022RPP,Hardouin2022NatComm,Rorai2022PRL,Adkins2022science,Cavaiola2022PoF,Puggioni2022PRE,Henshaw2023PRF,Mukherjee2023NatPhys} For example, \citet{Wu2000PRL} reported that due to the existence of large-scale motions, mass transport is enhanced.  \citet{Ishikawa2011PRL} experimentally measured how energy was transported from the individual cell to the larger mesoscale in 3D so that the energy is dissipated mainly in the large wave number regime. According to their estimation, the energy dissipated by mesoscale eddies is much smaller than the energy required for bacterial swimming.  Based on 3D experimental observation, \citet{Dunkel2013PRL} proposed a minimal fourth-order vector model to reproduce the main statistical characteristics of self-sustained turbulence in concentrated bacterial suspensions. 
Using a hydrodynamic model, \citet{Doostmohammadi2017NC} studied the onset of mesoscale turbulence in the channel. They reported
that the transition to mesoscale turbulence is
governed by the dimensionless activity number $A=h/\ell_{\alpha}$, the ratio of channel height $h$ to the characteristic activity-induced length scale $\ell_{\alpha}=\sqrt{K/\zeta}$ ($\zeta$ is intrinsic activity and $K$ is the orientational elasticity of the
nematic fluid): when $A\ge A_{\mathrm{cr}}$, similar to the classical hydrodynamic turbulence,\citep{Avila2011Science} the active puff grows into mesoscale turbulence.
\citet{Bratanov2015PNAS} reported numerically an inverse energy cascade in the Fourier representation for the continuum model. \citep{Wensink2012PNAS,Dunkel2013PRL}
Power law behavior of velocity energy  spectra was observed even in the absence of an inertial
range, but the spectral exponents may depend on the choice of parameters.
\citet{Slomka2017PNAS} found that the inverse energy cascade could be triggered in 3D if the mirror-symmetry is broken.  
\citet{Linkmann2019PRL} performed a systematic study of the parameters of the hydrodynamic model of dense microswimmer suspensions. They found a phase transition between bacterial turbulence (i.e., spatiotemporal chaos) and hydrodynamic turbulence (i.e., large-scale coherent structures). 
 However, it is still challenging to understand the intermittency characteristics for such a low Reynolds number turbulence-like system. \citep{Carenza2020EPL,Alert2022ARCMP,Aranson2022RPP}  \citet{Wensink2012PNAS} checked the high-order
 structure function for the experimental velocity field; due to the scale mixture of this method,\citep{Huang2010PRE} no scaling behavior is observed.  \citet{Mukherjee2023NatPhys} performed a series numerical simulation using the hydrodynamic model proposed by \citet{Wensink2012PNAS}. They reported a non-Gaussian fluctuation of velocity increments,  a signature of the intermittency when the level of activity beyond a critical value. They also proposed an asymptotic  energy spectrum $E(k)\propto k^{-3/2}$.
 By applying the Hilbert-Huang Transform to the experimental velocity field of 2D bacterial turbulence, when the scale mixture is overcome in a joint amplitude-frequency domain, a clear power law behavior is then obtained in the range $0.15 \lesssim  k\tilde{R} \lesssim 0.5$ with an intermittency parameter $\mu\simeq 0.26$ (see definition in Eqs.\,\eqref{eq:K62} and \eqref{eq:generalized_K62}) is reported.\citep{Qiu2016PRE} 
 To avoid the problem of the scale mixture in the physical domain,  \citet{Wang2017PRE} proposed a coordinate-free approach, namely the 
 streamline-based intrinsic flow structure analysis, to extract the flow structure in a natural-like coordinate. When applied to the 
 experimental velocity field, the power law behavior is confirmed in the range $2\lesssim \ell /\tilde{R}\lesssim 10$, corresponding to a range of wave numbers $0.1\lesssim k \tilde{R}\lesssim 0.5$, with a comparable intermittency parameter $\mu\simeq 0.20$. They also show 
evidence of the inverse energy cascade using the Filter-Space Technique (FST), and of the lognormal statistics of the energy dissipation rate  and enstrophy fields. These results confirm the existence of the inverse energy cascade even in the absence of an inertial range; moreover, contrary to conventional 2D hydrodynamic turbulence,  the intermittent correction coexists for this special active dynamic system.

In this work, cascades and associated lognormal statistics of the 2D bacterial turbulence will be further explored by analyzing the experimental velocity field. Experimental evidence is presented for the inverse energy cascade and the forward enstrophy cascade.  
The Kolmogorov lognormal scaling formula is also discussed.

\section{Experimental Data and Filter Space Technique}
\subsection{Experiment  Velocity Field of 2D Bacterial Turbulence}
The velocity vectors of the 2D bacterial turbulence analyzed here are experimental results provided by Goldstein. \citep{Wensink2012PNAS} Here, we briefly recall the main control parameters in a microfluidic chamber. \textit{Bacillus subtilis} is used for the experiment with an individual body length in the range $0.5\lesssim R/\tilde{R}\lesssim 1.5$ (i.e., $\tilde{R}\simeq 5\,\si{\mu m}$ for the mean body length) and a mean aspect ratio $a\simeq 6.3$, the ratio between body length $R$ and body diameter $d$. The vertical height $H_c$ of the microfluidic chamber is less than or equal to the length of the individual body to ensure 2D flow. The volume fill fraction is $\phi=84\%$ to ensure the turbulent phase of the flow.\citep{Wensink2012PNAS} The particle image velocimetry measurement area is $217\,\si{\mu m}\times 217\,\si{\mu m}$ with image resolution $700\,\si{pix}\times 700\,\si{pix}$ and frame rate $40\,\si{Hz}$. The final velocity field has $84\times 84$ vectors and a total of 1441 snapshots.  In total, there are 10,167,696 velocity vectors, which provide good statistical convergence. \citep{Wensink2012PNAS}  As shown by \citet{Wang2017PRE}, the flow field is smooth enough to safely calculate its spatial gradient, e.g., vertical vorticity $\omega_z$, energy dissipation rate $\epsilon$. Therefore, vorticity-associated enstrophy flux, and multifractal analysis of both the energy dissipation rate $\epsilon$ and the enstrophy $\Omega=\omega_z^2$ are considered.

\subsection{Filter Space Technique}

To determine both the direction and the strength of the cascade, one has to extract the scale-to-scale flux that characterizes the energy or other physical quantities exchanged between scales above $\ell>r$ and below $\ell<r$. \citep{Frisch1995} There exist mainly three methodologies to fulfill this job: third-order longitudinal
structure-function,\citep{Cerbus2017PoF,Zhang2021SA} spectral representation,\citep{Biferale2012PRLb,Pouquet2017PoF,Barjona2017PoF,Fathali2019PoF} and Filter-Space-Technique,\citep{Chen2006PRL,Ni2014PoF,Zhou2016JFM,Fang2017PoF,Wang2017PRE} respectively.  Although the first approach has been widely used in the turbulence community, it requires \textit{a priori} knowledge of the balance of the external force and the dissipation of the system. Thus, it is not suitable here to extract the scale-dependent energy flux information.  The second one provides a global representation of the energy flux in the Fourier space without providing local dynamic information. \citep{Alexakis2018PR,Zhou2021PR} The last method was first proposed in the turbulence community to \textit{a posteriori} extract the scale-to-scale flux
of the velocity field and preserve local dynamic information,\citep{Leonard1975AG,Eyink1995JSP,Eyink2009PoFa}   thus attracting more and more attention in not only fluid dynamics but also geophysical fluid dynamics. \citep{Chen2006PRL,Ni2014PoF,Zhou2016JFM,Wang2017PRE,Aluie2018JPO,Dong2020JFM,Loose2023JPO,Zhang2024FMS}

Taking the 2D velocity field, e.g. $\mathbf{u}(\mathbf{x},t)=[u_x(x,y,t),u_y(x,y,t)]$, as an example, its coarse-grained lower-pass field with spatial scale $r$ is defined as, 
\begin{equation}
    \mathbf{u}^{[r]}(\mathbf{x},t)=\mathbf{u}(\mathbf{x},t) \ast G_r(\mathbf{x},t)=\int \mathbf{u}(\mathbf{x}+\mathbf{x}',t) G_r(\mathbf{x}',t)
\mathrm{d} \mathbf{x}',
\end{equation}
where $\ast$ is the convolution, $G_r(\mathbf{x})$ is a filter kernel, and $r$ is the spatial scale. Due to its good low-pass property in Fourier space, the Gaussian kernel, i.e., $G_r(\mathbf{x})\propto \exp(- \mathbf{x}
^2/2r^2)$, is then often taken as the filter kernel. \citep{Boffetta2012ARFM} The scale-to-scale energy flux can be derived from the incompressible Navier-Stokes equations as follows,\citep{Eyink1995JSP} 
\begin{equation}
    \Pi^{[r]}(\mathbf{x},t)=-\sum_{i,j=1,2}\left[ (u_iu_j)^{[r]} -(u_i^{[r]}u_j^{[r]})
 \right]\frac{\partial u^{[r]}_i}{\partial x_j},
    \label{eq:FST}
\end{equation}
where $u_1=u_x,\,u_2=u_y$ is the velocity component, $x_1=x,\,x_2=y$ are the spatial coordinates in the 2D plane, and $r$ is the radius of the Gaussian kernel.  A positive $\Pi^{[r]}>0$ indicates the energy transferred from large-scale motions with spatial scales $\ell> r$ to small-scale ones $\ell< r$, and vice
versa. Thus, it characterizes both the direction and intensity of the energy cascade. 
It can be generalized to other physical quantities, e.g,
$\omega_z$,
\begin{equation}
    \Pi_{\Omega}^{[r]}(\mathbf{x},t)=-\sum_{i=1,2}\left[ (u_i\omega_z)^{[r]} -(u_i^{[r]}\omega_z^{[r]})
 \right]\frac{\partial \omega_z^{[r]}}{\partial x_i},
    \label{eq:FST_phi}
\end{equation}
Its interpretation is the same as that of the kinetic energy. 
The FST method has also been experimentally proven to be robust with noisy or underresolution
data. \citep{Ni2014PoF}

The cascade associates deeply with the nonlinear interaction of the governing
equation, e.g., $\mathbf{u}\cdot \nabla \mathbf{u}$,  in the Navier-Stokes equations. However, additional nonlinear interactions have been introduced in the model equations proposed by several authors. \citep{Wensink2012PNAS,Shaebani2020NRP}  To simplify the problem, here only the advection term $\mathbf{u}\cdot
\nabla \mathbf{u}$ is taken into account, as other effects are difficult to determine by using the experimental data. \citep{Wang2017PRE}

\begin{figure*}[!htb]
\centering
 \includegraphics[width=0.85\linewidth,clip]{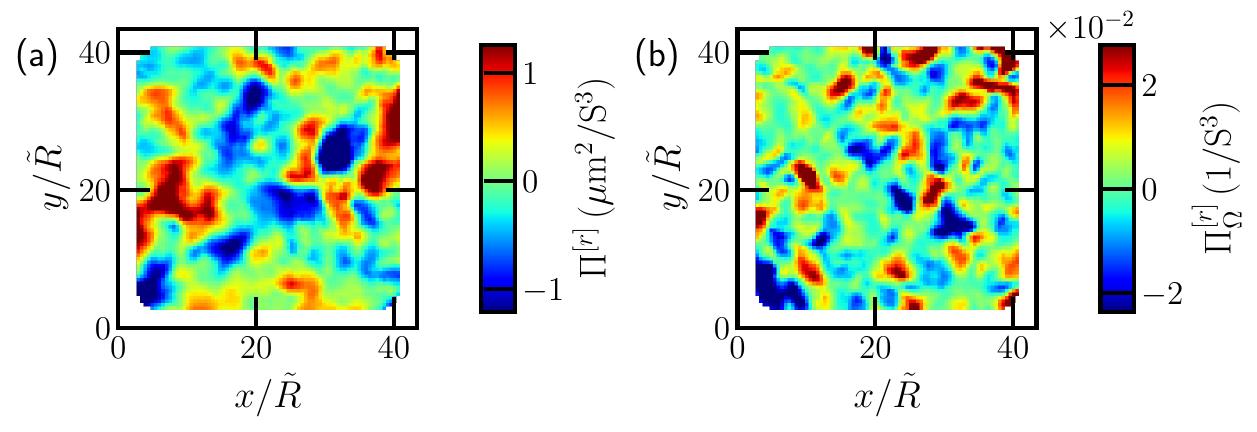}
  \caption{(Color online) A snapshot of (a) the energy flux  $\Pi^{[r]}$ 
and (b) enstrophy $\Pi_{\Omega}^{[r]}$  flux at scale $r/\tilde{R}=5$.
  }\label{fig:Snapshot}
\end{figure*}

Figure \ref{fig:Snapshot} shows an experimental example of scale-to-scale (a) energy flux $\Pi^{[r]}$  and (b) enstrophy
flux $\Pi_{\Omega}^{[r]}$  with $r/\tilde{R}=5$. Unlike the global Fourier spectral representation,\cite{Frisch1995,Bratanov2015PNAS}  the spatial
pattern has been well captured. For example, both the negative (inverse) and positive (forward) fluxes are local preserved. It is interesting to see that high-intensity negative fluxes are often accompanied by high-intensity positive fluxes, which is also observed for 3D homogeneous shear turbulence.\citep{Dong2020JFM}

\section{Results}

\subsection{Energy Cascade}

\begin{figure*}[!htb]
\centering
 \includegraphics[width=0.85\linewidth,clip]{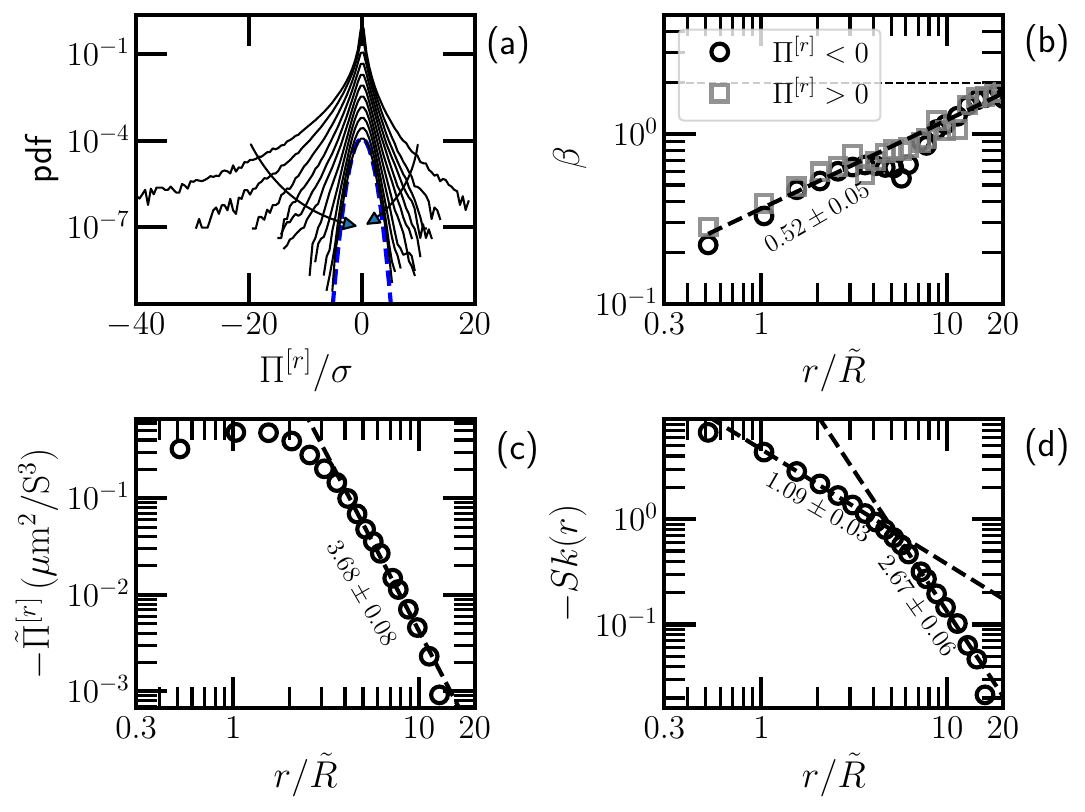}
  \caption{(Color online) (a) Measured  probability density function (pdf) of instantaneous 
  scale-to-scale energy flux $\Pi^{[r]}$ for scales in the range $0.5\lesssim r/\tilde{R}
\lesssim 20$, where the normal distribution is
illustrated as a dashed line. For display clarity, the pdf curves have been vertical shifted. (b) Stretched exponential distribution exponent $\beta$, where a power-law trend is demonstrated by a dashed line.
(c) Measured  mean energy flux 
  $\tilde{\Pi}^{[r]}$, where the dashed line is the power-law fit with a
scaling exponent $3.68\pm0.08$.  (d) The corresponding skewness factor, where the
dashed line indicates the power-law relation with scaling exponents $1.09\pm0.03$
and $2.67\pm0.06$.
  }\label{fig:Energy_Flux}
\end{figure*}

To address both the direction and intensity of cascades, the FST approach is applied to the experimental data of the 2D bacterial turbulence to retrieve scale-to-scale energy  and enstrophy  fluxes.  Note that the 2D velocity field has $84\times 84$ vectors and a spatial resolution
$2.58\,\si{\mu m}$, which is approximately half the mean body size $\tilde{R}$
(e.g, $5\,\si{\mu m}$).\citep{Wensink2012PNAS} Therefore, the spatial scale separation in these data
is approximately $40$ times. For the sake of good statistics, the largest scale is limited to $r/\tilde{R}\simeq 20$  in the following analysis. 
Figure\,\ref{fig:Energy_Flux}\,(a) shows the experimental probability density
function (pdf) of the energy flux $\Pi^{[r]}$ for scales in the range $0.5\lesssim r/\tilde{R}\lesssim 20$. It shows a negatively
skewed shape, indicating an inverse energy cascade.
With an increase in the scale $r$, it is from the stretched exponential distribution $p(x)\propto e^{-\vert x\vert^\beta}$ approaching the normal one (resp. $\beta\simeq 2$). \citep{Ching1991PRA}  The experimental $\beta$ through a nonlinear least squares  is shown in Fig.\,\ref{fig:Energy_Flux}\,(b). A power law increasing trend is observed with a scaling exponent $0.52$.
Note that the measured $\beta$ is slightly asymmetric for the left and right parts of the pdf and approaches the value 2 of the normal distribution. 
Moreover, the left part of the measured  $\beta$ deviates notably from the power law trend around the scale $r/\tilde{R}\simeq 5$. It follows again this power law trend when $r/\tilde{R}\gtrsim 8$.

Figure\,\ref{fig:Energy_Flux}\,(c) shows the mean energy flux $-\tilde{\Pi}^{[r]}$
in the log-log view.  Power law decaying is evident in the range $4\lesssim
r/\tilde{R} \lesssim 12$ with a scaling exponent $3.68\pm0.08$, suggesting that the
kinetic energy is strongly dissipated at all scale $r$, since these scale of motions are still below the fluid viscosity scale; see more discussion in section \ref{sec:dicusssion}. The scaling range agrees well with the one observed in Ref.\,\onlinecite{Wang2017PRE}, that is to be $2\lesssim r/\tilde{R}\lesssim 10$. To emphasize the asymmetry of pdfs, the skewness factor of each scale $r$ is calculated as follow,
 \begin{equation}
     Sk=\frac{\left\langle x^3 \right\rangle }{\left\langle x^2 \right\rangle^{3/2}},
 \end{equation}
where $\langle \cdot \rangle$ means the average of the ensemble, and $x$ is either the energy flux $\Pi^{[r]}$ or enstrophy flux $\Pi_{\Omega}^{[r]}$. The experimental $Sk(r)$ is shown in Fig.\,\ref{fig:Energy_Flux}\,(d). It is interesting to see a dual power law with scaling exponents, respectively, $1.09\pm 0.03$ of the range $1\lesssim r/\tilde{R} \lesssim 4.5$, and $2.67\pm 0.06$ of the range $5.5\lesssim
r/\tilde{R}\lesssim 14.5$. The separation scale $r/\tilde{R}\simeq 5$ agrees well with the one indicated by $\beta$, see Fig.\,\ref{fig:Energy_Flux}\,(b).

\subsection{Enstrophy Cascade}

\begin{figure*}[!htb]
\centering
 \includegraphics[width=0.85\linewidth,clip]{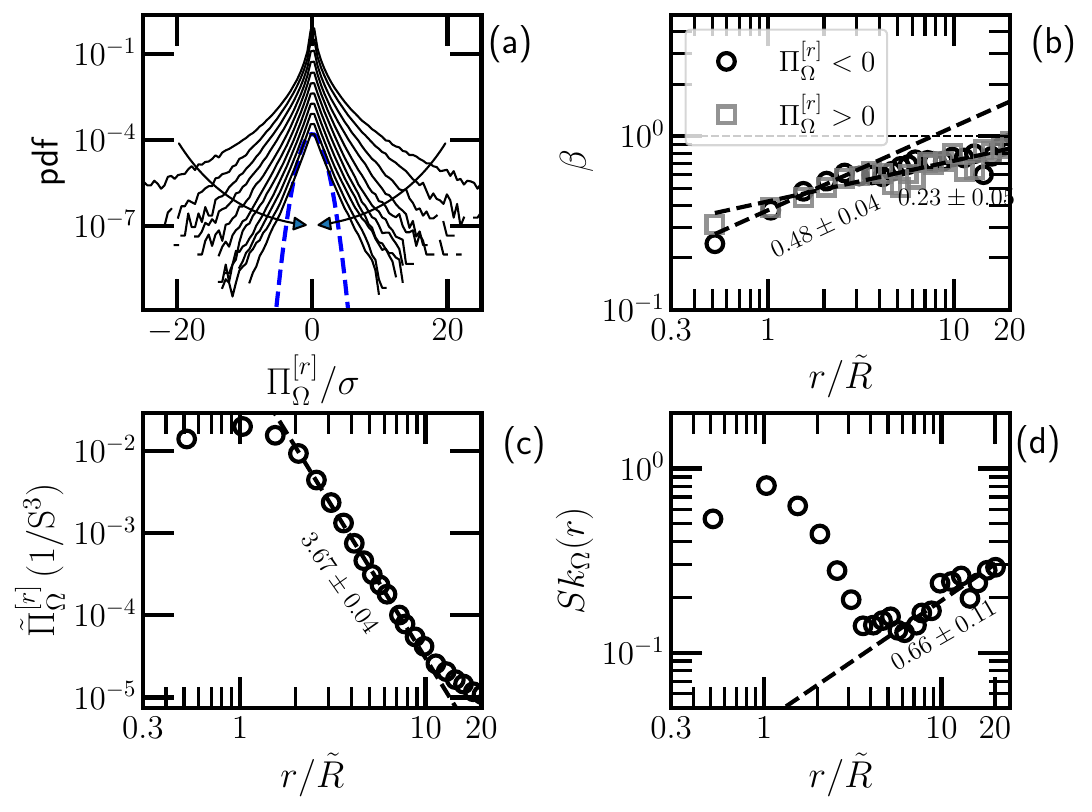}
  \caption{(Color online) (a) Measured pdf of the enstrophy flux $\Pi_{\Omega}^{[r]}$, where the normal distribution is
illustrated as a dashed line.  For display clarity, the pdf curves have been vertical shifted. (b) Stretched exponential distribution exponent $\beta$, where a power-law trend is demonstrated by a dashed line.
(c) Measured mean enstrophy flux $\tilde{\Pi}_{\Omega}^{[r]}$, where the
dashed line is the power-law fit with a scaling exponent $3.67\pm 0.04$.
(d) The corresponding skewness factor, where the dashed line indicates the power-law
scaling with a scaling exponent $0.66\pm0.11$. 
  }\label{fig:Enstrophy_Flux}
\end{figure*}

The experimental pdf of the enstrophy flux $\Pi_{\Omega}^{[r]}$ is shown in Fig.\,\ref{fig:Enstrophy_Flux}\,(a).
Unlike those of the energy flux, they seem to be symmetric for all scales.  With an increase in scales $r$, they are approaching the standard exponential distribution with $\beta\simeq 1$; see experimental $\beta$  in  Fig.\,\ref{fig:Enstrophy_Flux}\,(b). For the scales $r/\tilde{R}\lesssim 4$, experimental $\beta$ follows the power law increasing with a scaling exponent $0.48$, which is comparable to the one found for energy fluxes. Above $r/\tilde{R}\simeq 4$, the power law is less steep, showing a scaling exponent of value $0.23$.  The mean flux of enstrophy $\tilde{\Pi}_{\Omega}^{[r]}$ is positive with a decaying power law in the scale range $2\lesssim r/\tilde{R} \lesssim
8$ and a scaling exponent $3.67\pm 0.04$; see Fig.\,\ref{fig:Enstrophy_Flux}\,(c). It suggests that enstrophy might be injected into the system on all scales through other mechanisms; see more discussion in section \ref{sec:dis_cascade}.
The corresponding skewness factor suggests two regimes separated by a scale around $r/\tilde{R}\simeq 5$, where a power law increase is observed in the range $5.5\lesssim
r/\tilde{R} \lesssim 20$ with a scaling exponent $0.66\pm0.11$; see Fig.\ref{fig:Enstrophy_Flux}\,(d). Note that both the mean enstrophy flux and its skewness factor are positive for all scales, confirming the forward enstrophy cascade and a transition scale around $r/\tilde{R}\simeq 5$.

\subsection{Background Cascades}

\begin{figure*}[!htb]
\centering
 \includegraphics[width=0.85\linewidth,clip]{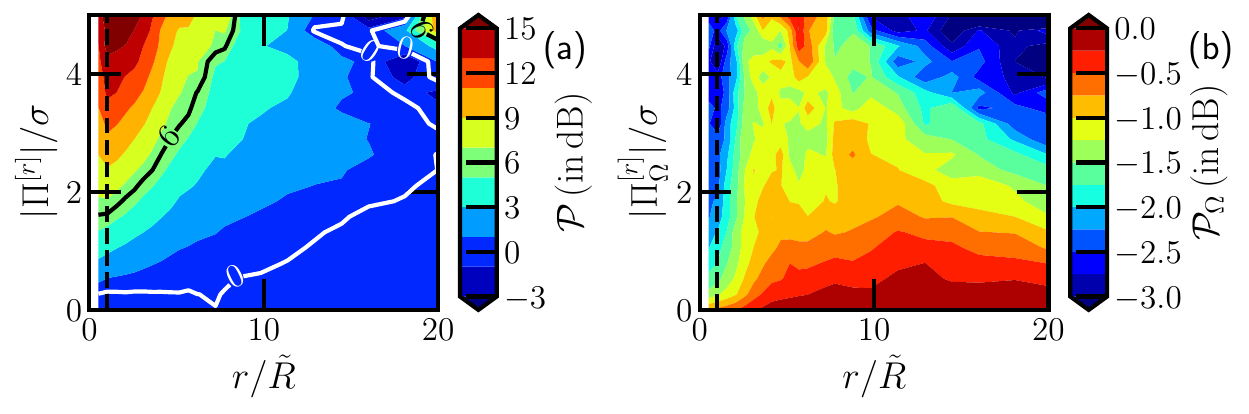}
  \caption{(Color online) Experiment negative versus positive pdf ratio $\mathcal{P}$
in dB. (a) $r$-dependent energy flux and (b) enstrophy flux, where the vertical
dash line indicates the body size of the bacteria.  The solid line in (a) indicates, respectively, 0\,dB and 6\,dB (i.e., twice).
  }\label{fig:PDF_ratio}
\end{figure*}

The part with the highest probability (resp. the core part of the pdf) represents the background motion of the flow field. \citep{Huang2011PoF} To better understand the contribution of different intensity events to the mean cascade,  a pdf ratio $\mathcal{P}$ in dB is defined as, 
\begin{equation}
\mathcal{P}(r,\vert \Pi^{[r] \vert})=20\times\log_{10} \frac{p(\Pi^{[r]})_{\Pi^{[r]}<0}}{p(\Pi^{[r]})_{\Pi^{[r]}>0}},
\end{equation}
where a positive value of $\mathcal{P}$ means that the inverse transfer is stronger than the forward transfer at a given intensity of events.  Figure \ref{fig:PDF_ratio}\,(a) shows the experimental $\mathcal{P}$ for the energy flux in the intensity range $0\lesssim \vert \Pi^{[r]} \vert/ \sigma
\lesssim 5 $, where $\sigma$ is its standard deviation. It is interesting to note that in the core part of the pdf, $\mathcal{P}$ is negative with a minimum value of approximately $-3$\,dB (that is roughly $0.7$ times), indicating that the background energy cascade of this active flow is still forward. In other words, the forward and inverse energy cascades coexist. \citep{Rorai2022PRL} With an increase in the intensity of the event, especially around body size, it is strongly positive, confirming that the energy is injected through the movement of bacteria. A maximum value $46$\,dB (approximately 200 times) is found for $r/\tilde{R}\simeq 1.5$ and $\Pi^{[r]}/\sigma\simeq
16.8$ (not shown here).
For the enstrophy case, the experimental $\mathcal{P}$ is nearly negative for all $r$ and $\Pi_{\Omega}^{[r]}$, indicating that the forward enstrophy cascade is dominant. It is interesting to note that the minimum value was also found to be around $-3$\,dB.

\subsection{Energy-Enstrophy Joint Cascades}\label{sec:EECascade}

 \begin{figure*}[!htb]
\centering
 \includegraphics[width=1\linewidth,clip]{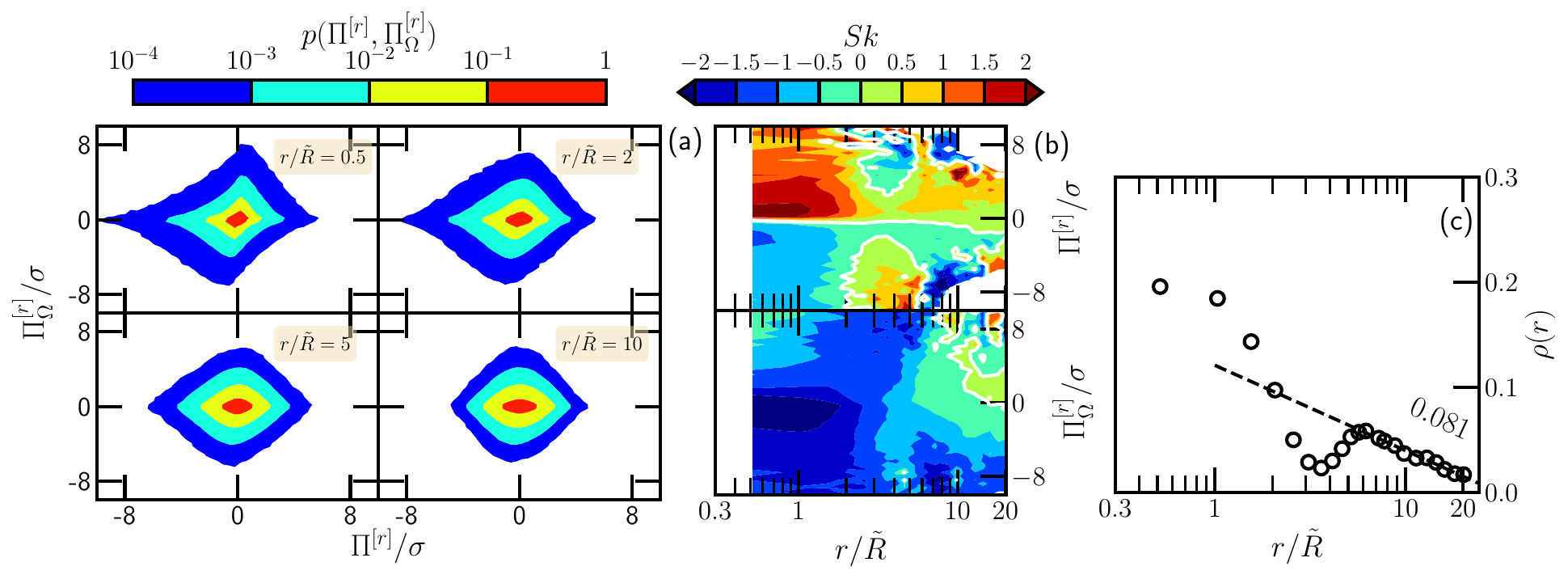}
  \caption{(Color online) (a) Experiment joint pdf $p(\Pi^{[r]},\Pi_{\Omega}^{[r]})$
for four typical spatial scales. (b) Conditional
skewness  $Sk(r,\Pi^{[r]})$ of enstrophy flux (top) and  $Sk(r,\Pi_{\Omega}^{[r]}
)$ of energy flux (bottom), where $Sk=0$ is indicated by a white line.  (c)
Cross-correlation coefficient $\rho(r)$ between $\Pi^{[r]}$ and $\Pi_{\Omega}^{[r]}$,
where the dashed solid line indicates  a log-law with a slope $0.081\pm0.004$
in the range $6\lesssim r/\tilde{R} \lesssim 20$.
  }\label{fig:Joint_PDF}
\end{figure*}

The energy and enstrophy cascades could be treated as a joint cascading process because they are dynamically related. \citep{Meneveau1990PRA,Nelkin1994,Nelkin1999PoF} Figure \ref{fig:Joint_PDF}\,(a) shows the measured joint pdf $p(\Pi^{[r]},\Pi_{\Omega}^{[r]})$ for scales in the range $0.5\lesssim r/\tilde{R} \lesssim 10$, in which the first four orders of magnitude are shown.  For small $r$, it is strongly asymmetric; with increasing scales $r$, it then approaches the up-down and right-left symmetry. To see more details of the asymmetry of the conditional pdf, the conditional skewness for $\Pi_{\Omega}^{[r]}$ and $\Pi^{[r]}$ are calculated, see Fig.\,\ref{fig:Joint_PDF}\,(b). The conditional
$Sk(r,\Pi_{\Omega}^{[r]})$ of the enstrophy flux (top) shows an antisymmetric relation with respect to $\Pi^{[r]}=0$. More precisely, for positive $\Pi^{[r]}$, except for special regions $3\lesssim r/\tilde{R} \lesssim 5$ and $2\lesssim \Pi^{[r]}/\sigma
\lesssim 8$, the conditional skewness factor is positive. For the case of energy flux, it is nearly negative for all scales $r$ and $\Pi_{\Omega}^{[r]}$. 

The experimental cross-correlation coefficient between $\Pi^{[r]}$ and
$\Pi_{\Omega}^{[r]}$ is shown in Fig.\,\ref{fig:Joint_PDF}\,(c). A log-law is evident
in the range $6\lesssim r/\tilde{R} \lesssim 20$ with a scaling exponent $0.081\pm0.04$.
All of the above results confirm the existence of an inverse energy cascade and a
forward enstrophy cascade below the fluid viscosity scale and a finite scaling
behavior with a scaling exponent  $n\frac{2}{3}$ (e.g., $n=0,2,3$, see Tab.\,\ref{tab:scaling}), \citep{Wensink2012PNAS,Qiu2016PRE,Wang2017PRE}
and indicate a separation scale around $r/\tilde{R}\simeq 5$.

\subsection{Kolmogorov Lognormal Statistics}

\begin{figure*}[!htb]
\centering
  \includegraphics[width=1\linewidth,clip]{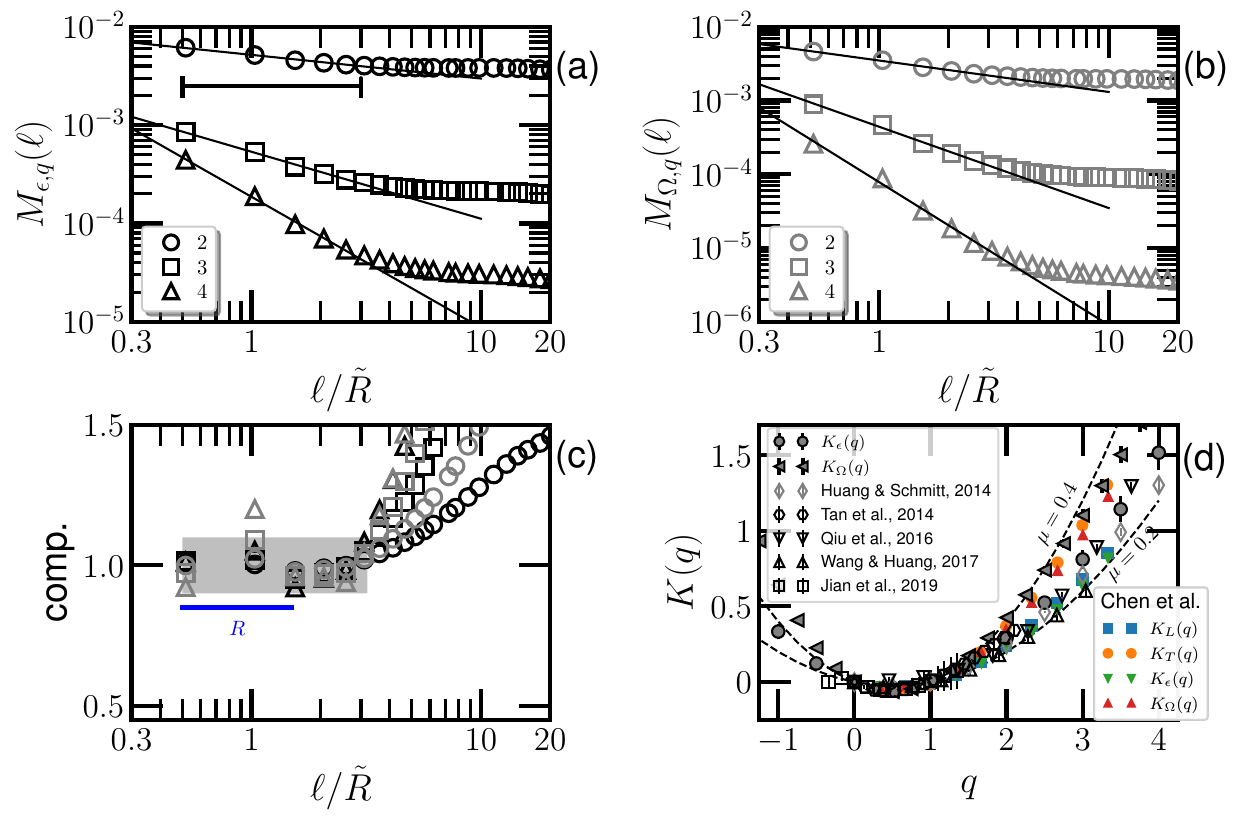}
  \caption{ (a) Measured high-order moments $M_{\epsilon,q}(\ell)$, where a power-law behavior is evident
on the range $0.5\lesssim \ell/\tilde{R}\lesssim 3$. (b) The same as (a) but for $M_{\Omega,q}(\ell)$.
(c) The compensated curves to emphasize the experimental power-law behavior. 
(d) Experimental $K_{\epsilon}(q)$ (\ding{108}, $\mu_{\epsilon}=0.26\pm0.01$ ) and $K_{\Omega}(q)$ ($\blacktriangleleft$, $\mu_{\Omega}=0.36\pm0.02$).
 The $K(q)$ calculated for the same dataset through Eq.\,\eqref{eq:generalized_K62}
using $\zeta(q)$ provided in Ref.\,\onlinecite{Qiu2016PRE} ($\bigtriangledown$, $\mu\simeq 0.26$) and in
Ref.\,\onlinecite{Wang2017PRE} ($\triangle$, $\mu\simeq 0.20$) are also shown.
The directly calculated $K(q)$  for the 3D Lagrangian turbulence $\Diamond$ ($\mu\simeq 0.23$) provided in Ref.\,\onlinecite{Huang2014JFM}; indirectly ones provided in Ref.\,\onlinecite{Tan2014PoF} for the forward cascade in the 2D turbulence ($\hexagon$, $\mu\simeq 0.31$); 
Ref.\,\onlinecite{Chen1997PRL} for both longitudinal (blue $\color{blue}\blacksquare$, $\mu\simeq 0.23$) and transverse (orange \textcolor{orange}{\ding{108}}, $\mu\simeq 0.35$) scaling, directly measure $K_{\epsilon}(q)$ (green $\color{green}{\blacktriangledown}$, $\mu_{\epsilon}\simeq 0.21$)   and $K_{\Omega}(q)$ (red \textcolor{red}{\ding{115}}, $\mu_{\Omega}\simeq 0.33$) in Ref.\,\onlinecite{Chen1997PRLb},  and the lithosphere deformation $\square$ ($\mu\simeq 0.30$) in Ref.\,\onlinecite{Jian2019PRE} are also shown. The intermittency parameter $\mu$ is provided by the least squares of either Eqs.\,\eqref{eq:K62} or \eqref{eq:generalized_K62} in the range $0\leq q \leq 4$. 
  }\label{fig:Kq}
\end{figure*}

In his 1941 theory of 3D HIT, Kolmogorov considered the mean energy dissipation rate as the key parameter.\citep{Kolmogorov1941,Frisch1995} As Landau pointed out, the energy dissipation rate $\epsilon$ (see the definition below) should be varied in both space and time, leading to a non-universal spectrum for different flows. \citep{Landau1987} In 1962, 20 years after his 1941 theory, Kolmogorov proposed his famous refined theory of turbulence, in which the lognormal statistics of the energy dissipation field are used to reconcile the highly intermittent distribution of the energy dissipation rate in space.\citep{Kolmogorov1962,Frisch1995,Yao2024PRL} The energy dissipation rate of the velocity field is defined as, 
\begin{equation}
\epsilon(
\mathbf{x},t)=2\nu S_{ij}S_{ij},\, S_{ij}=\frac{1}{2}\left( \frac{\partial u_i}{\partial
x_j}+\frac{\partial u_j}{\partial x_i} \right),
\end{equation}
where $\nu$ is the fluid viscosity.  A coarse-grained energy dissipation
rate $\epsilon_{\ell}$ is defined as 
\begin{equation}
\epsilon_{\ell}(\mathbf{x},t)=\frac{1}{D}\int_{\vert \mathbf{x}'\vert \leq \ell}\epsilon(\mathbf{x}+\mathbf{x}',t)
\mathrm{d}\mathbf{x}',
\end{equation}
where, following the convention, the radius $\ell$ is the characteristic spatial scale;\citep{Taylor1915eddy,Taylor1918dissipation,Kolmogorov1962} $D=\pi \ell ^2$  for the 2D case and $D=\frac{4}{3}\pi \ell ^3$ for the
3D case. The high-order moment of $\epsilon_{\ell}$ is expected to follow the power law decaying if $\epsilon_{\ell}$ satisfies the lognormal distribution, that is, 
\begin{equation}
M_{\epsilon,q}(\ell)=\langle  \epsilon_{\ell}^q\rangle \propto \ell^{-K_{\epsilon}(q)}, \, K_{\epsilon}(q)=\frac{\mu_{\epsilon}}{2}\left(q^2-q\right),\label{eq:K62}
\end{equation}
where $\mu_{\epsilon}$ is the so-called intermittency parameter. A higher value of $\mu_{\epsilon}$,
a more intermittent energy dissipation field.

An experimental test has been performed in Ref.\,\onlinecite{Wang2017PRE} to confirm the validation of the lognormal distribution of both $\epsilon_{\ell}$ and $\Omega_{\ell}$ for all $\ell$ (figure not reproduced here).
Figure\,\ref{fig:Kq}\,(a) shows the high-order moment $M_{\epsilon,q}(\ell)$ of the
coarse-grained energy dissipation rate $\epsilon_{\ell}$.  The power law decaying is observed in a finite range $0.5\lesssim
\ell/\tilde{R}\lesssim 3$. 
Note that, for example in the statistical analysis,\citep{Chou1963Book,Huang2017PRE}   if one treats the diameter as the spatial scale, the scaling range is then in the range $1\lesssim 2\ell /\tilde{R}\lesssim 6$.
The experimental $M_{\Omega,q}(\ell)$ for the enstrophy is shown in Fig.\,\ref{fig:Kq}\,(b), where power law decaying is evident in the same range of scales as that of the energy dissipation rate. The compensated curves using the fitted parameters are shown in Fig.\,\ref{fig:Kq}\,(c) to highlight the experimental power law behavior.
Note that the finite scale range of power law scaling has been recognized by several different methodologies and quantities,  e.g., less than one decade in Figs.\,\ref{fig:Energy_Flux}, \ref{fig:Enstrophy_Flux}, Fig.\,4 in Ref.\,\onlinecite{Wensink2012PNAS}, Figs.\,4, 5 and 6 in Ref.\,\onlinecite{Qiu2016PRE}, and Fig.\,5 in Ref.\,\onlinecite{Wang2017PRE}. 
There are several reasons to limit the scaling range since the flow motion is below the fluid viscosity scale; see the comment in Ref.\,\onlinecite{Qiu2016PRE} and the section \ref{sec:FRSB}.

The direct measurement of $K_{\epsilon}(q)$ and $K_{\Omega}(q)$ are shown in Fig.\,\ref{fig:Kq}\,(d), yield  intermittency parameters  $\mu_{\epsilon}\simeq 0.26$ and $\mu_{\Omega}\simeq 0.36$, confirming the validation of Eq.\,\eqref{eq:K62}. These intermittency parameters are comparable to those for high Reynolds number 3D hydrodynamic turbulence.\citep{Sreenivasan1993PoF,Frisch1995,Chen1997PRL,Chen1997PRLb,Huang2014JFM} For example,  the best fit of intermittency parameters $K_{\epsilon}(q)$ and $K_{\Omega}(q)$ in Ref.\,\onlinecite{Chen1997PRLb} for the 3D homogeneous and isotropic turbulence are respectively $\mu_{\epsilon}\simeq0.21$ and $\mu_{\Omega}=0.33$. These values suggest a more intermittent enstrophy field than energy dissipation and have been further recognized as a finite Reynolds number effect. \citep{Nelkin1994,Nelkin1999PoF,Yeung2012JFM,Iyer2019NJP} For comparison, the directed measured $K_{\epsilon}(q)$ for the 3D Lagrangian turbulence with an intermittency parameter $\mu_{\epsilon}\simeq 0.23$ provided in Ref.\,\onlinecite{Huang2014JFM} is also shown in Fig.\,\ref{fig:Kq}\,(d). The experimental $K(q)$ for various flows show a good agreement, indicating that the intermittency parameter $\mu$ might be universal; see more discussions in section \ref{sec:universality}.

\subsection{Kolmogorov Lognormal Scaling Relation}
The scaling of the energy dissipation field is deeply related to Kolmogorov's 1962 refined turbulence theory that the intermittency feature of the velocity field stems from the wild distribution of the former quantity.\citep{Kolmogorov1962,Frisch1995} Taking the 3D hydrodynamic turbulence as an example, the scaling exponent $K_{\epsilon}(q)$ of the energy dissipation field and $\zeta_E(q)$ of the Eulerian structure-function of the velocity field can be related as follows,
 \begin{equation}
 \zeta_E(q)=\frac{q}{3}-\frac{\mu_{\epsilon}}{18}\left( q^2-3q\right),\label{eq:K62E}
 \end{equation} 
Its Lagrangian counterpart is written as follows,
 \begin{equation}
 \zeta_L(q)=\frac{q}{2}-\frac{\mu_\epsilon}{8}\left( q^2-2q\right),\label{eq:K62L}
 \end{equation} 
A generalization of the above formula for an arbitrary Hurst number $h$ is written as follows,
\begin{equation} 
\zeta(q)=qh-K(qh),\,K(q)= \frac{\mu}{2}(q^2-q),\label{eq:generalized_K62}
\end{equation}
where $h$ is the Hurst number provided by either theoretical considerations (e.g., $1/3$ for the Eulerian
velocity or $1/2$ for the Lagrangian velocity), or determined by the least squares fit of $\zeta(q)$. This lognormal scaling relation has been verified for the 3D HIT in  either Eulerian and Lagrangian frames.\citep{Stolovitzky1992PRL,Stolovitzky1994,Chen1997PRL,Chen1997PRLb,Yu2010PRL,Benzi2009PRE,Huang2014JFM} For example, the indirect measure $\mu$ for the longitudinal and transverse structure-function scaling in Ref.\,\onlinecite{Chen1997PRL} are $\mu\simeq 0.23$ and $\mu\simeq 0.35$; they agree well with the direct measure values $\mu_{\epsilon}\simeq 0.21$ and $\mu_{\Omega}\simeq 0.33$ in Ref.\,\onlinecite{Chen1997PRLb}.

Therefore, indirect measurements of $K(q)$ are also calculated using experimental $\zeta(q)$ through the relation $K(q)=q-\zeta(q/h)$. The scaling exponent $\zeta(q)$ of this bacterial velocity field has been obtained by two different methods: Hilbert-Huang Transform \citep{Qiu2016PRE} and streamline-based analysis. \citep{Wang2017PRE} The indirect estimation of $K(q)$ is also shown in Fig.\,\ref{fig:Kq}\,(d) as $\bigtriangledown$ and $\triangle$, where the Hurst number $h$ obtained by the nonlinear least squares fit. Interestingly, the direct and indirect measured $K(q)$ ($\mu\simeq 0.26$  of Ref.\,\onlinecite{Qiu2016PRE} and  $\mu\simeq 0.20$ of Ref.\,\onlinecite{Wang2017PRE}) agree well with
each other, confirming the validation of the aforementioned Kolmogorov lognormal scaling relationship. This suggests that the intermittency correction observed in 2D bacterial turbulence could originate from the energy dissipation field, which deserves more careful study in the future; see more discussion in Section \ref{sec:universality}.

For comparison, $K(q)$ calculated either directly or indirectly from other flow systems are also shown: (1)  the 3D Lagrangian HIT ($\Diamond$) in Ref.\,\onlinecite{Huang2014JFM}, Eulerian longitudinal (blue $\textcolor{blue}\blacksquare$) and transverse (orange \textcolor{orange}{\ding{108}}) provided in Ref.\,\onlinecite{Chen1997PRL}, $K_{\epsilon}(q)$ (green $\color{green}{\blacktriangledown}$) and
$K_{\Omega}(q)$ (red \textcolor{red}{\ding{115}}) provided in Ref.\,\onlinecite{Chen1997PRLb}; (2) the forward cascade in the 2D turbulence ($\hexagon$, $\mu\simeq 0.31$) in Ref.\,\onlinecite{Tan2014PoF};
(3)  the deformation of the lithosphere ($\square$, $\mu\simeq 0.30$) in Ref.\,\onlinecite{Jian2019PRE}.  Despite very different
methodologies and systems, the experimental $K(q)$ agree very well with
each other, suggesting a possible universal intermittency correction, e.g. $0.2\lesssim \mu \lesssim 0.4$. \citep{Sreenivasan1993PoF,Frisch1995}
Note that one of the shortcomings of the lognormal model is that it may not be realizable since it predicts unlimited velocities; see more comments in Ref.\,\onlinecite{Frisch1995}.
Other multifractal models \citep{Parisi1985,Benzi1984,Schertzer1987,She1994PRL,Dubrulle1994PRL} could also well fit the scaling exponents in Fig.\,\ref{fig:Kq}\,(d) or scaling relation Eq.\,\eqref{eq:generalized_K62}, in which other analytical forms for $K(q)$ may be involved. To keep things as simple as possible, we do not discuss other models here.

\subsection{Joint Multifractal Measures}

\begin{figure*}[!htb]
\centering
  \includegraphics[width=0.85\linewidth,clip]{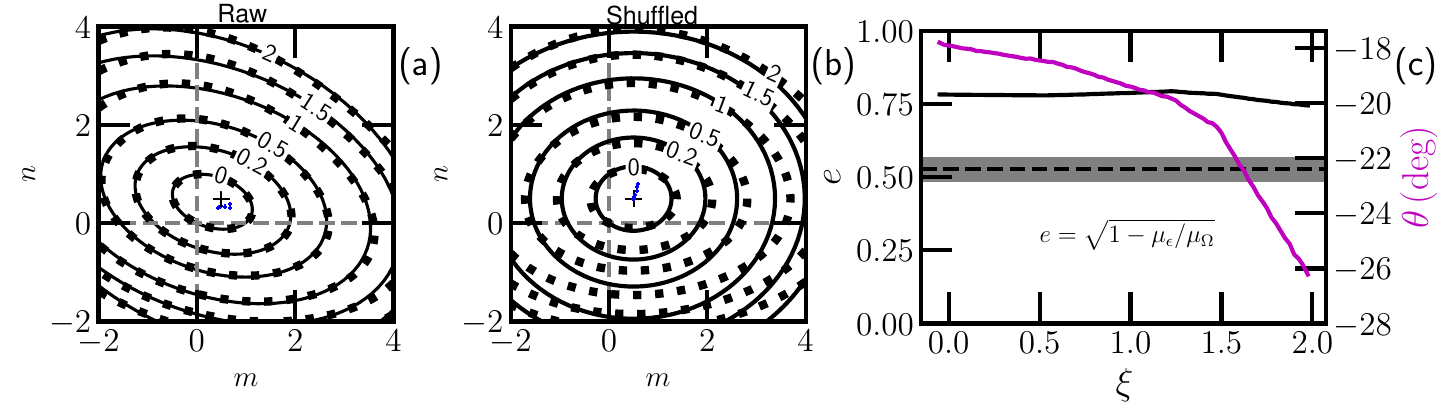}
  \caption{ (a)  Contour lines of measured $\xi(m,n)$ (dashed line), where the solid line is an ellipse fit.  (b) The decoupling test $\xi(m,n)$ (dashed line), where the analytical form provided by Eq.\,\eqref{eq:ellipse} with experimental $\mu_{\epsilon}=0.26$ and
  $\mu_{\Omega}=0.36$ is shown as a solid line. (c) The eccentricity of measured $\xi(m,n)$ (solid line), where the value of $e=\sqrt{1-\mu_\epsilon/\mu_\Omega}$ (i.e., $0.53\pm0.04$) is shown as a dashed line.  The magenta solid line 
  is the fitted inclined angle $\theta$ (counterclockwise) of the measured $\xi(m,n)$. The measured ellipse center is illustrated as a blue cross in (a) and (b).
  }\label{fig:joint_MFM}
\end{figure*}

In Section \ref{sec:EECascade}, the joint distribution of energy and enstrophy fluxes is examined, which confirms the existence of the joint energy-enstrophy cascade. This joint cascade can be further characterized by the so-called joint multifractal measures using mixed high-order moments of the coarse-grained energy dissipation rate and enstrophy,\citep{Meneveau1990PRA,Consolini2023FF} which is written as,
\begin{equation}
M_{m,n}(\ell)=\langle \epsilon_{\ell}^m \Omega_{\ell}^n \rangle \propto \ell^{-\xi(m,n)},
\end{equation}
where $\xi(m,n)$ is the scaling exponent that satisfies $\xi(m,0)=K_{\epsilon}(m)$ and $\xi(0,n)=K_{\Omega}(n)$. When the multifractal fields $\epsilon$ and $\Omega$ are independent of each other, one has,
\begin{equation}
\xi(m,n)=K_{\epsilon}(m)+K_{\Omega}(n),
\end{equation}
By substituting Eq.\,\eqref{eq:K62} into the above equation,  it yields an ellipse model for the joint scaling exponents $\xi(m,n)$,\citep{Meneveau1990PRA}
\begin{equation}
\frac{\mu_{\epsilon}}{2}\left( m -\frac{1}{2}\right)^2+\frac{\mu_{\Omega}}{2}\left( n -\frac{1}{2}\right)^2=\frac{1}{8}(\mu_{\epsilon}+\mu_{\Omega})+\xi(m,n),\label{eq:ellipse}
\end{equation}
To test the above relation, one has to destroy the correlation between $\Omega_{\ell}(\mathbf{x},t)$ and $\epsilon_{\ell}(\mathbf{x},t)$: they are shuffled in both space and time to decouple with each other.  The high-order moments $M_{m,n}(\ell)$ are then calculated, respectively, with and without shuffling in the range $-2\leq m,n\leq 4$. The experimental scaling exponents $\xi(m,n)$ are then estimated in the range $0.5\lesssim \ell /\tilde{R}\lesssim 3$.
Figure \ref{fig:joint_MFM}\,(a) and (b) show the contour lines of the measured $\xi(m,n)$ ($\blacksquare$) without and with shuffling.
Note that in Fig.\,\ref{fig:joint_MFM}\,(b), contour lines provided by Eq.\,\eqref{eq:ellipse} with experimental values of $\mu_{\epsilon}=0.26$ and $\mu_{\Omega}=0.36$ are shown as a solid line without further adjustment. The decoupled $\xi(m,n)$ agrees well with the analytical ellipse model in a wide range of $-0.08\lesssim \xi(m,n)\lesssim 1$. 
However, the direct measured $\xi(m,n)$ seems to follow the ellipse shape with an inclined angle.
 Therefore, the eccentricity $e=\sqrt{1-b^2/a^2}$ ($a$ and $b$ are the major and minor axes of the ellipse) and the inclined angle $\theta$ (counterclockwise) are extracted. The measured $e$ is found to be a constant value of $e=0.78\pm0.01$, which is significant above the value of $0.53\pm0.04$ (i.e., $\sqrt{1-\mu_\epsilon/\mu_\Omega}$) provided by Eq.\,\eqref{eq:ellipse}. However, the inclined angle $\theta$ decreases with $\xi$ in the range $-26^{\circ}\lesssim \theta \lesssim -18^{\circ}$. More precisely, two regimes separated by $\xi\simeq 1.5$ with different slopes (e.g., $-1.87\pm0.03$ and $-10.67\pm0.06$) can be identified.

\section{Discussion}\label{sec:dicusssion}

\subsection{Scale-Dependent Cascade and Dissipation}\label{sec:dis_cascade}

\begin{figure*}[!htb]
\centering
  \includegraphics[width=0.85\linewidth,clip]{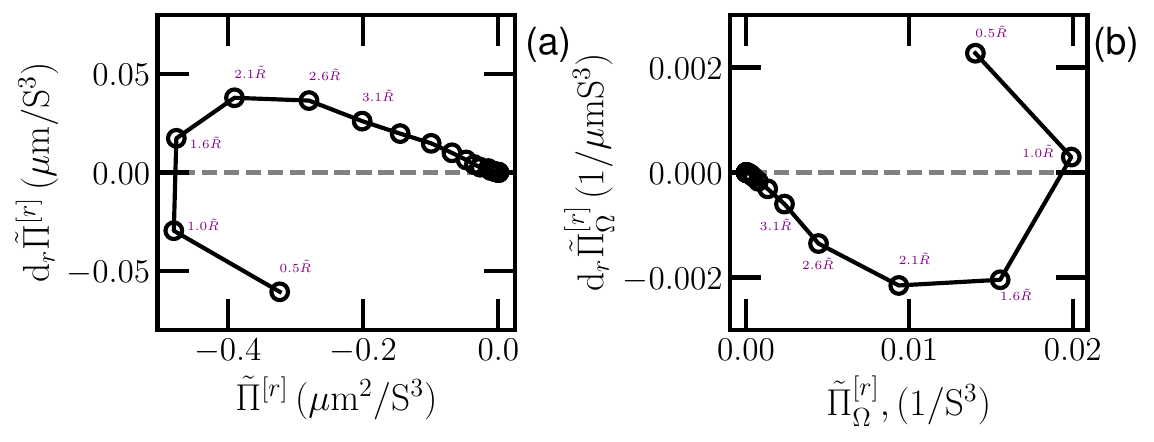}
  \caption{(a) Phase diagram of the energy flux $\tilde{\Pi}^{[r]}$ versus $\mathrm{d}_r\tilde{\Pi}^{[r]}$. (b) the enstrophy flux $\tilde{\Pi}_{\Omega}^{[r]}$ versus $\mathrm{d}_r\tilde{\Pi}_{\Omega}^{[r]}$.
  }\label{fig:B_dissipation}
\end{figure*}

\begin{figure*}[!htb]
\centering
  \includegraphics[width=0.85\linewidth,clip]{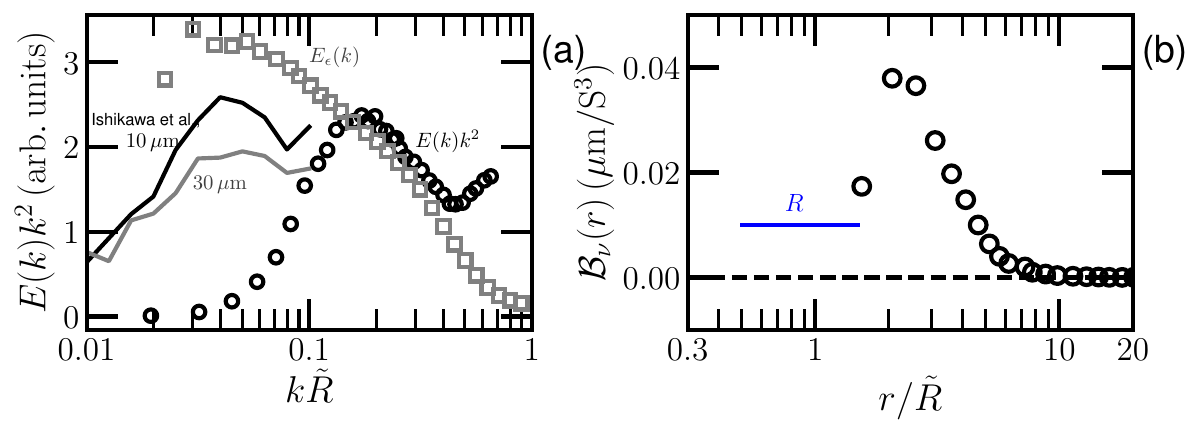}
  \caption{(a) Normalized experimental dissipation spectrum $E(k)k^2$ ($\ocircle$), where a peak is found to be a $k\tilde{R}\simeq 0.2$, corresponding to a spatial scale $\ell\simeq 5\tilde{R}$. For comparison, the Fourier power spectrum of the energy dissipation rate is shown as $\square$, and the dissipation spectrum in Ref.\,\onlinecite{Ishikawa2011PRL} shown as solid line. For display clarity, the solid lines have been vertically shifted. (b) The experimental energy dissipation density function $\mathcal{B}_{\nu}(r)$.
  }\label{fig:dissipation_spectrum}
\end{figure*}

Note that in the global balance Eq.\,\eqref{eq:balance}, the energy flux $\tilde{\Pi}^{[r]}$ is not involved, as it is only exchanged between different scales internally. Assuming the spatial homogeneity and temporal stationary, the local energy balance on scale $r$ is written as,\citep{Frisch1995,Zhang2024FMS,Qin2024JOES}
\begin{equation}
    \mathrm{d}_r\tilde{\Pi}^{[r]}=\mathcal{B}_{\nu}(r) -\mathcal{E}_{\mathrm{in}}(r),\label{eq:local_B} 
\end{equation}
Here, a negative value of $\mathrm{d}_r \tilde{\Pi}^{[r]}$ indicates an injection of energy into the loop of the cascade and vice verse. With different combinations of these three terms, different pictures of cascades can be realized.  For example, far from the injection scale $L$ and the dissipation scale $\eta$, a scale-independent energy flux $\tilde{\Pi}^{[r]}$ is required, since $\mathrm{d}_r\tilde{\Pi}^{[r]}=0$. This is the inertial range for both 3D and 2D HITs.\citep{Kraichnan1974JFM}
Figure \ref{fig:B_dissipation}\,(a) shows the experimental phase diagram of $\tilde{\Pi}^{[r]}$ versus $\mathrm{d}_r\tilde{\Pi}(r)$ through a finite center difference. It confirms that the kinetic energy is injected via the bacterial body size (i.e.,  $\mathrm{d}_r\tilde{\Pi}^{[r]}<0$ when $r\simeq \tilde{R}$), and is dissipated out the system on all scales (i.e., $\mathrm{d}_r\tilde{\Pi}^{[r]}>0$) that above the injection scale $\tilde{R}$. In other words, the kinetic energy is indeed dissipated on all scales.\citep{Ishikawa2011PRL} 
This can be also confirmed by the the so-called dissipation spectrum,\citep{Monin1971} i.e., $E_{\epsilon}(k)=E(k)k^2$,\footnote{The dissipation spectrum $E_{\epsilon}(k)$ is not the Fourier power spectrum of the energy dissipation rate $\epsilon$. It comes from the mathematical relation $\epsilon=\int E(k)k^2\mathrm{d} k$, where $E(k)$ is the kinetic energy spectrum.} where $E(k)$ is the kinetic energy spectrum of the velocity. Figure \ref{fig:dissipation_spectrum}\,(a) reproduces the normalized experimental curve $E(k)k^2$, where $E(k)$ is adopted from Ref.\,\onlinecite{Wensink2012PNAS}. A peak of the dissipation spectrum is evident at $k\tilde{R}\simeq 0.2$, corresponding to
a spatial scale $\ell\simeq 5\tilde{R}$. The normalized Fourier power spectrum of the energy dissipation rate is also shown as $\square$. A log-law decaying is evident with a slope $2.20\pm0.03$ on the range $0.07\lesssim k\tilde{R} \lesssim 0.2$, corresponding to a spatial scale range $5\lesssim \ell/\tilde{R} \lesssim 14$. For comparison, the dissipation spectrum in the 3D bacterial turbulence provided in Ref.\,\onlinecite{Ishikawa2011PRL} is also reproduced as solid lines, where two heights above the bottom of their experiments are $10\,\mathrm{\mu m}$ and $30\,\mathrm{\mu m}$. As mentioned above, when the scale $r$ is above the size of the body of the bacteria, one can treat $\mathrm{d}_r\tilde{\Pi}(r)$ as $\mathcal{B}_{\nu}(r)$; see Fig.\,\ref{fig:dissipation_spectrum}\,(b). All these results suggest and confirm that the kinetic energy is dissipated on all scales $r$.

However, the bacterial turbulence is very different from that of classical HITs as follows,
\begin{itemize}
    \item[1)] Movement is below the fluid viscosity scale $\eta$ when the viscosity reduction effect is not considered; \citep{Sokolov2009PRL,Rafai2010PRL,Lopez2015PRL}
    \item[2)] The scaling range, if it exists, is limited by the body size of bacteria $R$ and the fluid viscosity scale $\eta$;\citep{Wang2017PRE}
    \item[3)] Because of 1), the kinetic energy is dissipated at all scales $r$, e.g., $\mathcal{B}_{\nu}(r)>0$. \citep{Ishikawa2011PRL}
\end{itemize}
Based on the above observations, the following picture of cascade can be drawn. The movement of bacteria is mainly balanced by the viscosity of the fluid immediately at its size $\tilde{R}$.\citep{Ishikawa2011PRL} Only a small portion of the energy is injected into the loop of the energy cascade. However, due to the presence of fluid viscosity, it rapidly dissipates on all scales $r$.  Therefore, there is no constant flux as expected for the classical inertial turbulence.
A summary of this cascade picture of the bacterial turbulence referred to above is shown in Fig.\,\ref{fig:cascade}\,(c).

Concerning the enstrophy cascade, it is more complex: for example, it is forward, but its intensity decreases with $r$; see Fig.\,\ref{fig:Enstrophy_Flux}\,(c). Negative $\mathrm{d}_r\tilde{\Pi}_{\Omega}^{[r]}$ indicates a relation $\mathcal{E}_{\Omega,\mathrm{in}}(r)>\mathcal{B}_{\Omega}(r)>0$. In other words, the enstrophy is injected into the system on all scales $r$. One possible reason is due to the local alignment between the vorticity gradient and the space transport of the vorticity, that is, $(u_j\omega_z)^{[r]} -(u_j^{[r]}\omega_z^{[r]})$, at different scales,\citep{Chen2006PRL,Buaria2020PRF} or other nonlinear interactions associated with the enstrophy cascade rather than the advection term $\mathbf{u}\cdot \nabla \mathbf{u}$, but not captured here for the experimental data, which deserves more careful study in the future using proper model equations.

\subsection{Finite Range of Scaling Behaviors}\label{sec:FRSB}

\begin{table}[!htb]
\caption{Scaling ranges and intermittency parameters $\mu$ observed in the experimental velocity of 2D bacterial turbulence provided in Ref.\,\onlinecite{Wensink2012PNAS}. The intermittency parameters measured for 3D HIT are $\mu_\epsilon\simeq 0.21$ and $\mu_\Omega\simeq 0.33$. \citep{Chen1997PRLb}}\label{tab:scaling}
\begin{tabular}{ |p{3cm}|p{3cm}|p{3cm}|p{3cm} | p{3cm}|  }
 \hline
 \hline
 Method/Quantity & Scaling Range  & Scaling exponents & Intermittency $\mu$  & Ref. \\[1ex]
 \hline
Fourier  & $2\lesssim r /\tilde{R} \lesssim 5$   &$2.68 (2\frac{2}{3})\pm0.03$  & NA & \citet{Wensink2012PNAS} \\[1ex]
  \hline
Hilbert &  $2\lesssim r /\tilde{R} \lesssim 6$   & $0.91\pm0.02$  &$0.26\pm0.01$ & \citet{Qiu2016PRE}\\[1ex]
\hline
 Streamline & $2\lesssim r /\tilde{R}\lesssim 10$ & $0.76  \pm0.01$ & $0.20\pm0.01$ &Wang \& Huang\citep{Wang2017PRE}\\[1ex]
 \hline
 $\beta$ of $\Pi^{[r]}$   & $0.5\lesssim r /\tilde{R}\lesssim 20$ \newline $r /\tilde{R} \simeq 5 $ &  $0.52\pm0.05$ & NA & Fig.\,\ref{fig:Energy_Flux}\,(b) \\[1ex]
 \hline
 $\tilde{\Pi}^{[r]}$ & $4\lesssim r /\tilde{R}\lesssim 12$ &  $3.68(3\frac{2}{3})\pm0.08$ & NA & Fig.\,\ref{fig:Energy_Flux}\,(c)\\[1ex]
\hline
 $Sk(r)$ of $\Pi^{[r]}$ & $1\lesssim r /\tilde{R}\lesssim 4.5$ \newline $5.5\lesssim r /\tilde{R}\lesssim 14.5$ &  $1.09\pm0.03$ \newline $2.67(2\frac{2}{3})\pm0.06$ & NA & Fig.\,\ref{fig:Energy_Flux}\,(d)\\[1ex]
 \hline
 $\beta$ of $\Pi_{\Omega}^{[r]}$   & $0.5\lesssim r /\tilde{R}\lesssim 4$ \newline $4\lesssim r /\tilde{R}\lesssim 20$ &  $0.48\pm0.04$ \newline $0.23\pm0.05$ & NA & Fig.\,\ref{fig:Enstrophy_Flux}\,(b) \\[1ex]
 \hline
 $\tilde{\Pi}_{\Omega}^{[r]}$ & $2\lesssim r /\tilde{R}\lesssim 8$ &  $3.67(3\frac{2}{3})\pm0.04$ & NA & Fig.\,\ref{fig:Enstrophy_Flux}\,(c)\\[1ex]
\hline
 $Sk(r)$ of $\Pi_{\Omega}^{[r]}$ & $5.5\lesssim r /\tilde{R}\lesssim 20$  &  $0.66(\frac{2}{3})\pm0.11$  & NA & Fig.\,\ref{fig:Enstrophy_Flux}\,(d)\\[1ex]
 \hline
$\rho(r)$ & $6\lesssim r /\tilde{R}\lesssim 20$  &  $0.081\pm0.004$  & NA & Fig.\,\ref{fig:Joint_PDF}\,(c)\\[1ex]
 \hline
 $M_{\epsilon,q}(\ell)$ & $0.5\lesssim \ell /\tilde{R}\lesssim 3$  & NA  & $\mu_{\epsilon}:0.26\pm0.01$ & Fig.\,\ref{fig:Kq}\,(a)\\[1ex]
 \hline
$M_{\Omega,q}(\ell)$ & $0.5\lesssim \ell /\tilde{R}\lesssim 3$  & NA  & $\mu_{\Omega}:\,0.36\pm0.02$ & Fig.\,\ref{fig:Kq}\,(b)\\[1ex]
 \hline 
\end{tabular}
\end{table}

Experimentally, scale separation is limited by the measurement technique; and by the scale ratio of the fluid viscosity scale $\eta$ and the mean bacterial body size $\tilde{R}$. Therefore, the scale range of the scaling behavior is short, e.g., less than one order magnitude of scales.  The scaling range and the corresponding scaling exponent or intermittency parameter $\mu$ are summarized in Tab.\,\ref{tab:scaling}, in which all scales are in the physical domain. We note that several scaling exponents are of the form $n\frac{2}{3}$, which is also indicated in Tab.\,\ref{tab:scaling} if their fractional parts are close to $2/3$.  For example, the scaling exponent of the Fourier power spectrum of the velocity field is found to be $2.68$,\citep{Wensink2012PNAS}  which is in the form $2\frac{2}{3}$. It suggests an experimental Hurst number $h=5/6$ through the scaling relation $\beta=1+2h$.\citep{Frisch1995,Schmitt2016Book}
A characteristic scale $r/\tilde{R}\simeq 5$ is observed for several quantities; for example, the Fourier and Hilbert spectrum of velocity,\citep{Wensink2012PNAS,Qiu2016PRE} energy flux $\tilde{\Pi}^{[r]}$ and its skewness factor, etc., to name a few. According to the convention, in the coarse-graining analysis, the spatial scale $\ell$ is the radius of the circle, while in statistical analysis, the spatial scale is often considered as the diameter. When the latter definition is taken, the scaling range is $1\lesssim 2\ell /\tilde{R}\lesssim 6$.
Moreover,  as noted by \citet{Kraichnan1974JFM}, different approaches over a scaling range interval do not necessarily make it into an inertial range quantity, e.g., energy dissipation rate $\epsilon$.

\subsection{Universality of Intermittency Corrections}\label{sec:universality}

Here, we quote directly the words by \citet{Kolmogorov1985} that published in 1985: ``\textit{Moreover, I soon understood that there was little hope of developing a pure, closed theory, and because of the absence of such a theory the investigation must be based on hypotheses obtained in processing experimental data. It was also important to have collaborators capable of combining theoretical and experimental research work}."   Following his spirit, data analysis should be performed for different  flow systems to pursue a data-inspired explanation or theory.  
When the experimental data are analyzed, the scaling curve $K(q)$ and the associated intermittency parameter $\mu$ are recovered in the lognormal framework, where the dependence of the Hurst number $h$ is excluded. Thus, the universality of the intermittency correction could be verified for various flows. Here, we show that the intermittency parameter $\mu$, estimated either directly or indirectly, are indeed comparable, at least in four quite different flow systems: classical 2D and 3D HITs, bacterial turbulence, lithosphere deformation, etc., to name a few. In other words, the strength of the intermittency correction might be universal. An elegant and more rigorous theoretical argument should be proposed to take into account this experimental observation.

Note that, unlike the classical 2D HIT, the intermittency correction and inverse energy cascade are simultaneously observed for the 2D bacterial turbulence. This is because the cascade of bacterial turbulence is below the viscosity scale $\eta$, where the fluid viscosity plays an important role. For example, kinetic energy is rapidly dissipated at all scales $r$ without the existence of a constant energy flux $\tilde{\Pi}^{[r]}$.
This scenario is very different from the inertial range observed in the 2D or 3D HITs, where the influence of the fluid viscosity can be ignored, so a constant energy flux $\tilde{\Pi}^{[r]}$ through scales $r$ is expected;
and for the 2D case, a non-intermittent scaling has been observed for the inverse cascade.\citep{Tan2014PoF}
For this active flow system, the scaling behavior could depend on the type of bacteria; and for the model, it could depend on the
choice of parameters. For example, bacterial turbulence emerges only when its concentration is within a range of values.\citep{Wensink2012PNAS} This is partially due to the fact that the value of concentration determines how far the cascade can go, since the kinetic energy is mainly balanced immediately at its body size:\citep{Ishikawa2011PRL} a low concentration means a low energy injection rate $\mathcal{E}_{\mathrm{in}}(r)$, the flow structure generated by an individual bacterium will soon be smoothed out by the fluid viscosity. Therefore, the inverse energy cascade, if it exists, is too short to be detected by the classical methodology.\citep{Qiu2016PRE} A systematic analysis of different parameters experimentally or numerically should be performed in the framework considered here, which is beyond the topic of this work.

Finally, we would like to make a remark on the low Reynolds number turbulence-like flows. As mentioned above, turbulence is believed to be one of the most important features of high Reynolds number flows, in which the viscosity force can be ignored if one explains the Reynolds number as the ratio between the inertial force and viscosity force. In fact, one may rewrite the Reynolds number as follows,
\begin{equation}
\mathrm{Re}=\frac{L}{\nu/u},
\end{equation}
It can be explained as the spatial scale ratio between the largest scale structure and the viscosity structure. In other words, the Reynolds number is one of those parameters that characterize scale separation of the systems. In the current system, spatial scale separation is roughly estimated at least as $\mathcal{O}(10^2)$. 

\section{Conclusions}

In summary, the cascade of the bacterial turbulence and the associated lognormal statistics are examined in this work. Our
data analysis confirms the inverse energy cascade and the forward enstrophy cascade below the fluid viscosity scale. Because of the presence of the inverse energy cascade, large-scale coherent structures are then generated through the hydrodynamic interaction spontaneously. However, the experimental energy flux decays rapidly with the increase of the spatial scale $r$ due to the strong influence of fluid viscosity; no constant energy flux is observed, since the kinetic energy is dissipated on all scales $r$. The degree to which the inverse energy is transferred depends not only on the viscosity of the fluid but also on the injection rate of the  kinetic energy, e.g., the concentration of bacteria. It is also evident that the background energy cascade is still forward. Moreover, the experimental pdfs of both energy and enstrophy fluxes can be described well by the stretched exponential distribution. For the former, with an increase of the spatial scale, it is approaching the standard normal distribution;  for the latter, it is approaching the standard exponential distribution. 
Regarding the cascade of enstrophy, the experimental results suggest an injection on almost all scales $r$. One possible reason is that the local alignment between the vorticity gradient and the space
transport of the vorticity, or other nonlinear interactions associated with the enstrophy cascade rather than the advection term are not captured in the experimental data.

Concerning the energy dissipation rate and enstrophy fields, the power law behavior of their high-order moments of coarse-grained multifractal fields are observed as expected. The corresponding scaling exponents $K(q)$ are well described by the lognormal formula with the intermittency parameters $\mu$ comparable with those of the 2D and 3D HITs. In addition, the lognormal scaling relation that connects the scaling behavior of the velocity field and the energy dissipation rate  through Eq.\,\eqref{eq:generalized_K62} are verified. In analogy to
Kolmogorov's 1962 refined theory of turbulence, this observation suggests that the intermittency correction in the velocity field may originate from the multifractal nature of the energy dissipation field. Unlike the classical 2D HIT, the inverse energy cascade and intermittency coexist for this special active dynamical system; see the summarized cascade picture in Fig.\,\ref{fig:cascade}\,(c) and discussions in Section\,\ref{sec:dis_cascade}.

The coupling between the energy and enstrophy cascades is characterized by the joint analysis of their fluxes and multifractal measures. The correlation on different scales $r$ is evident. The joint scaling exponent $\xi(m,n)$ of their multifractal measures can be well fitted by the ellipse formula of lognormal statistics.  Compared to the decoupling test, a higher eccentricity $e(\xi)$ with an inclined angle $\theta(\xi)$ is observed. They can be treated as a signature of the coupling between energy and enstrophy cascades. Our results show that the lognormal statistics is a relevant analogy to the framework of Kolmogorov's 1962 refined theory of 3D HIT. 
Models of active flows should reproduce not only the large-scale coherent structure, but all detailed scaling relations presented in this work, which is beyond the topic of this work.

Finally, the intermittency parameter $\mu$ might be universal for quite different flow systems.  For example, a comparable intermittency parameter $\mu_{\epsilon}$ in the range $0.21 \sim 0.23$ has been found for 3D HITs,\citep{Chen1997PRL,Chen1997PRLb,Huang2014JFM} $\mu\simeq 0.31$ for 2D HIT,\citep{Tan2014PoF}
and $\mu\simeq 0.30$ for the lithosphere deformation with extremely low Reynolds number, 
for example, $\mathrm{Re}\simeq \mathcal{O}(10^{-24})$. \citep{Jian2019PRE} More data analysis is needed for various turbulent flows or turbulence-like systems to extract their intermittency parameters to see whether the intermittency index within the lognormal framework is universal or not.

\begin{acknowledgments}
This work is sponsored by the National Natural Science Foundation (Nos. U22A20579). 
The author acknowledges Prof. R.E. Goldstein
at Cambridge University, UK for providing us with the experiment data.  
 I thank Professor Fran\c{c}ois G Schmitt, Enrico Calzavarini, Luca Biferale and Lipo Wang for the stimulating discussion on this topic.
 The reproduced data in Figs.\,\ref{fig:Kq}\,(d) and \ref{fig:dissipation_spectrum}\,(a) are extracted by using an open source software WebPlotDigitizer.\citep{Rohatgi2022}
 I would like to dedicate this work to the 100th anniversary (1922-2022) of Richardson's cascade and to the 120th anniversary of the birth of Professor Peiyua Chou  (August 28, 1902- November 24, 1993), the father of computational modeling.
\end{acknowledgments}

\section*{AUTHOR DECLARATIONS}
\section*{Conflict of Interest}
The authors have no conflicts to disclose.

\section*{Author Contributions}
Yongxiang Huang: Conceptualization (equal); Data curation (equal); Formal analysis (equal); Funding acquisition (equal); Investigation (equal); Methodology (equal); Project administration (equal); Resources (equal); Validation (equal); Writing - original draft (equal); Writing - review \& editing (equal).

\section{DATA AVAILABILITY}
The data that support the findings of this study are available
at \href{http://damtp.cam.ac.uk/user/gold/datarequests.html}{http://damtp.cam.ac.uk}.
A copy of the source code for the present analysis is available at
 {\href{https://github.com/lanlankai}{https://github.com/lanlankai}}. 





 

%

\end{document}